      [1999/12/01 v1.4c Il Nuovo Cimento]
\documentclass{cimento}

\usepackage{epsfig}
\usepackage{cite}

\title{Top quark pair production with two jets at next-to-leading order}
\author{Ma\l gorzata Worek\from{ins:x}\thanks{Presented at the 
3rd International Workshop on Top Quark Physics, Top 2010, 
Bruges, Belgium,
May 31 - June 4,  2010.}}
\instlist{\inst{ins:x} 
 Fachbereich C, Bergische Universit\"at Wuppertal, D-42097
      Wuppertal,Germany}
\PACSes{\PACSit{12.38.Bx}{Perturbative calculations}
\PACSit{14.65.Ha}{Top quarks}}

\begin{document}

\maketitle

\begin{abstract}
A report on the recent next-to-leading order QCD calculations to 
$t\bar{t}b\bar{b}$ and $t\bar{t}jj$ at 
the CERN Large Hardon Collider is given.  The elements 
of the calculation are  briefly summarized and results for integrated 
and differential cross sections are presented. 
\end{abstract}

\section{Introduction}

Even though first results for next-to-leading order (NLO) QCD corrections to
heavy quark production were presented in the late 80's and early 90's
\cite{Nason:1987xz,Beenakker:1988bq,Nason:1989zy,Beenakker:1990maa} the topic
of higher order corrections to $t\bar{t}$ is still very active and far from
complete. Our present level of understanding is very well summarized in
experimental and theoretical reviews, see  {\it e.g.} 
\cite{Cacciari:2008zb,Bernreuther:2008ju,Incandela:2009pf}.
Recent progress in NLO
\cite{Melnikov:2009dn,Czakon:2008ii}  and next-to-next-to leading order (NNLO)
\cite{Czakon:2007ej,Czakon:2007wk,Korner:2008bn,Czakon:2008zk,Bonciani:2008az,
Anastasiou:2008vd,Kniehl:2008fd,Bonciani:2009nb,Beneke:2009ye}
calculations, as well as next-to-next-to-leading-log resummations (NNLL)
\cite{Czakon:2008cx,Beneke:2009rj,Czakon:2009zw}  for inclusive $t\bar{t}$
hadroproduction is truly astonishing.
The list for the  more exclusive channels is just as impressive: NLO QCD
corrections have been calculated for the $t\bar{t}H$ signal
\cite{Beenakker:2001rj,Reina:2001bc,Reina:2001sf,Beenakker:2002nc,Dawson:2002tg,
Dawson:2003zu}, where the Higgs boson has been treated as a stable
particle. Most recently  the factorisable QCD corrections to this process
have been presented \cite{Binoth:2010ra}, where higher order corrections  to
both production and decay of the Higgs boson into a  $b\bar{b} $ pair have been
calculated. 
Moreover, NLO QCD corrections to a variety of $2 \to 3$ backgrounds processes
$t\bar{t}j$ \cite{Dittmaier:2007wz,Dittmaier:2008uj, Melnikov:2010iu},
$t\bar{t}Z$ \cite{Lazopoulos:2008de} and  $t\bar{t}\gamma$ \cite{PengFei:2009ph}
have been obtained. Last but not least, the NLO QCD corrections to    
the $2 \to 4$ backgrounds processes such as 
$t\bar{t}b\bar{b}$  \cite{Bredenstein:2008zb,Bredenstein:2009aj,
  Bevilacqua:2009zn,Bredenstein:2010rs}  and $t\bar{t}jj$
\cite{Bevilacqua:2010ve} have also recently been completed.

Both processes $pp \rightarrow t\bar{t}b\bar{b}$ and 
$pp \rightarrow t\bar{t}jj$ represent very important
background reactions to searches at the LHC, in particular to $t\bar{t}H$
production, where the Higgs boson decays into a $b\bar{b}$ pair. A successful
analysis of this  particular production channel requires the  knowledge of
direct  $t\bar{t}b\bar{b}$ and $t\bar{t}jj$ production at NLO in  QCD
\cite{Aad:2009wy}. In this contribution, a brief
report on these computations is given.

\section{Theoretical framework}

NLO QCD corrections have been calculated within the \textsc{Helac-Nlo} 
framework.  It consists of \textsc{Helac-Phegas} 
\cite{Kanaki:2000ey,Papadopoulos:2000tt,Cafarella:2007pc}, which has, 
on its own, already been extensively used and tested in phenomenological
studies see {\it e.g}
\cite{Gleisberg:2003bi,Papadopoulos:2005ky,Alwall:2007fs,Englert:2008tn}.
 \textsc{Helac-Phegas} is a multi-purpose, tree-level  event generator which 
is the only existing implementation of the algorithm based on Dyson-Schwinger 
equations. It can be used to efficiently obtain helicity amplitudes and total 
cross sections  for arbitrary multiparticle processes in the Standard Model. 
The program can generate all processes with 10 or more final state particles 
with full off-shell and finite width effects taking into account naturally 
both, spin and color correlations. The integration  over the fractions  
$x_1$ and $x_2$ of the  initial partons is done 
via \textsc{Parni} \cite{vanHameren:2007pt} .

Virtual corrections are obtained using the 
\textsc{Helac-1Loop} program \cite{vanHameren:2009dr}, based on the
Ossola-Papadopoulos-Pittau (OPP) 
reduction technique \cite{Ossola:2006us,Pittau:2010tk} and the reduction code 
\textsc{CutTools} \cite{Ossola:2007ax,Draggiotis:2009yb,Garzelli:2009is}. 
Moreover, 
\textsc{OneLOop} \cite{vanHameren:2009dr} library has been used for the
evaluation of the scalar integrals.
Reweighting techniques, and helicity and colour sampling methods 
are used  in order to optimize the performance of our system.
The OPP reduction at the integrand level takes advantage of the knowledge that
the final answer for one loop amplitudes can be expressed in terms of a basis 
of known $4-$, $3-$, $2-$ and  $1-$point scalar integrals: boxes, triangles, 
bubbles and tadpoles\footnote{Tadpole integrals are present only when there are 
internal  massive propagators.}:
\[
{\cal{A}} = \sum_i d_i I^4_i + \sum_i c_i I^3_i +   \sum_i b_i I^2_i + 
\sum_i a_i I^1_i + {\cal{R}}
\]
where ${\cal{R}}$ is the so called rational part and $d_i, b_i, c_i, a_i$ are 
coefficients which have to be derived. The OPP method aims at computing 
them directly avoiding any computationally intensive integral reduction. 

The OPP reduction is based on a representation of the numerator of amplitudes,
a polynomial in the integration momentum, in a basis of polynomials given by
products of the functions in the denominators. Clearly, the cancellation of
such terms with the actual denominators will lead to scalar functions with a
lower number of denominators. By virtue of the proof provided by the
Passarino-Veltman reduction \cite{Passarino:1978jh}, 
we will end up with a tower of four-point and
lower functions, as mentioned before. The determination of the decomposition
in the new basis proceeds recursively, by setting chosen denominators
on-shell. This is where the OPP method resembles generalized
unitarity \cite{Bern:1994zx,Bern:1994cg,Witten:2003nn,Britto:2004nc,
Anastasiou:2006jv,Anastasiou:2006gt,Bern:2007dw,Ellis:2007br,Giele:2008ve,
Ellis:2008ir}. For most recent applications see {\it e.g.} 
\cite{Berger:2009zg,Berger:2009ep,Berger:2010vm,Ellis:2009zw,
KeithEllis:2009bu,Melnikov:2009wh}. 
It is important to stress, that working around four dimensions, allows to
compute the numerator function in four dimensions. The difference to the
complete result is of order  $\epsilon$, and can therefore be
determined a posteriori in a simplified manner 
\cite{Draggiotis:2009yb,Garzelli:2009is}. Since the calculation of the
coefficients of the reduction requires the evaluation of the numerator
function for a given value of the loop momentum, the corresponding diagrams
can be thought of as tree level (all momenta are fixed) graphs. To complete the
analogy, one needs to chose a propagator and consider it as cut. At this point
the original amplitude for an $n$ particle process becomes a tree level
amplitude for an $n+2$ particle process. The advantage is that its value can be
obtained by a tree level automate such as \textsc{Helac-Phegas}. The bookkeeping
necessary for a practical implementation is managed by a new software,
\textsc{Helac-1Loop}.

The OPP method has already been successfully applied 
to a  large number of processes, apart from already mentioned 
$t\bar{t}b\bar{b}$, $t\bar{t}H \to t\bar{t}b\bar{b}$ and $t\bar{t}jj$ also to 
the production of three vector bosons, namely $Z Z Z$, $W^+ W^- Z$, 
$W^+ Z Z$  and $W^+ W^- W^+$ final states at the LHC 
\cite{Binoth:2008kt} and to the calculation of one-loop QED corrections 
to the hard-bremsstrahlung process $e^- e^+ \gamma$  at $e^- e^+$
colliders \cite{Actis:2009uq}. 
Recently   the OPP-approach has been implemented in the 
another framework called \textsc{Samurai} \cite{Mastrolia:2010nb}, 
together with an extention
 which accommodate an implementation of the generalized d-dimensional 
unitarity-cuts technique.

The singularities from soft or collinear parton emission are isolated via
dipole subtraction for NLO QCD calculations \cite{Catani:1996vz} using the
formulation for massive quarks \cite{Catani:2002hc} and for arbitrary
polarizations \cite{Czakon:2009ss}. After combining virtual and real
corrections, singularities connected to collinear configurations in the final
state as well as soft divergencies in the initial and final states cancel for
infrared-safe observables automatically. 
Singularities connected to collinear initial-state splittings are
removed via factorization by PDF redefinitions.   
Calculations are performed with the help of the \textsc{Helac-Dipoles} 
software \cite{Czakon:2009ss},
which is a complete and publicly available automatic implementation  of
Catani-Seymour dipole subtraction and consists of  phase space integration  of
subtracted real radiation and integrated dipoles in both massless and massive
cases.  The phase space restriction on the
contribution of the dipoles as originally proposed in
\cite{Nagy:1998bb,Nagy:2003tz} is also implemented. Two values of the
unphysical cutoff are always considered; $\alpha_{max}=1$, which corresponds to
the case when all dipoles are included, and $\alpha_{max}=0.01$. The 
independence of the final result  on this cutoff is explicitly checked in all 
our results, both for the integrated cross  section and for the differential 
distributions.  
Moreover, also in this part helicity sampling methods 
are used  in order to speed up the calculation.

The cancellation of divergences between the real and virtual corrections is 
always verified. In addition, the numerical precision of the latter has been
assured by using gauge invariance tests and use of quadruple precision.
Let us emphasise that all parts are calculated fully numerically
in a completely automatic manner.

Finally, the phase-space integration is performed with the multichannel 
Monte Carlo  generator \textsc{Phegas} \cite{Papadopoulos:2000tt} and 
\textsc{Kaleu} \cite{vanHameren:2010gg}.

\begin{table}
  \caption{Integrated cross section at LO and NLO for $t\bar{t}b\bar{b}$ 
  production at the LHC. The two NLO results refer to different values of 
  the dipole phase space cutoff $\alpha_{max}$. The scale choice is 
   $\mu_R=\mu_f=m_{top}$.}
  \label{tab:ttbb}
  \begin{tabular}{rcl}
    \hline
      $\sigma^{\rm{LO}}$ [fb]      & 
      $\sigma^{\rm{NLO}}_{\rm{\alpha_{max}=1}}$  [fb] &   
      $\sigma^{\rm{NLO}}_{\rm{\alpha_{max}=0.01}}$   [fb] \\
    \hline
      1489.2 $\pm$ 0.9    &  2642  $\pm$ 3   &  2636 $\pm$  3 \\
    \hline
  \end{tabular}
\end{table}
\begin{table}
  \caption{Integrated cross section at LO and NLO for $t\bar{t}jj$ production
  at the LHC. The two NLO results refer to different values of the dipole phase
  space cutoff $\alpha_{max}$. The scale choice is $\mu_R=\mu_f=m_{top}$.}
  \label{tab:ttjj}
  \begin{tabular}{rcl}
    \hline
      $\sigma^{\rm{LO}}$ [pb]      & 
      $\sigma^{\rm{NLO}}_{\rm{\alpha_{max}=1}}$  [pb] & 
      $\sigma^{\rm{NLO}}_{\rm{\alpha_{max}=0.01}}$   [pb]  \\
    \hline
      120.17 $\pm$ 0.08    &  106.95  $\pm$ 0.17   & 106.56  $\pm$ 0.31  \\
    \hline
  \end{tabular}
\end{table}
\begin{figure}
\begin{center}
\includegraphics[width=0.45\textwidth]{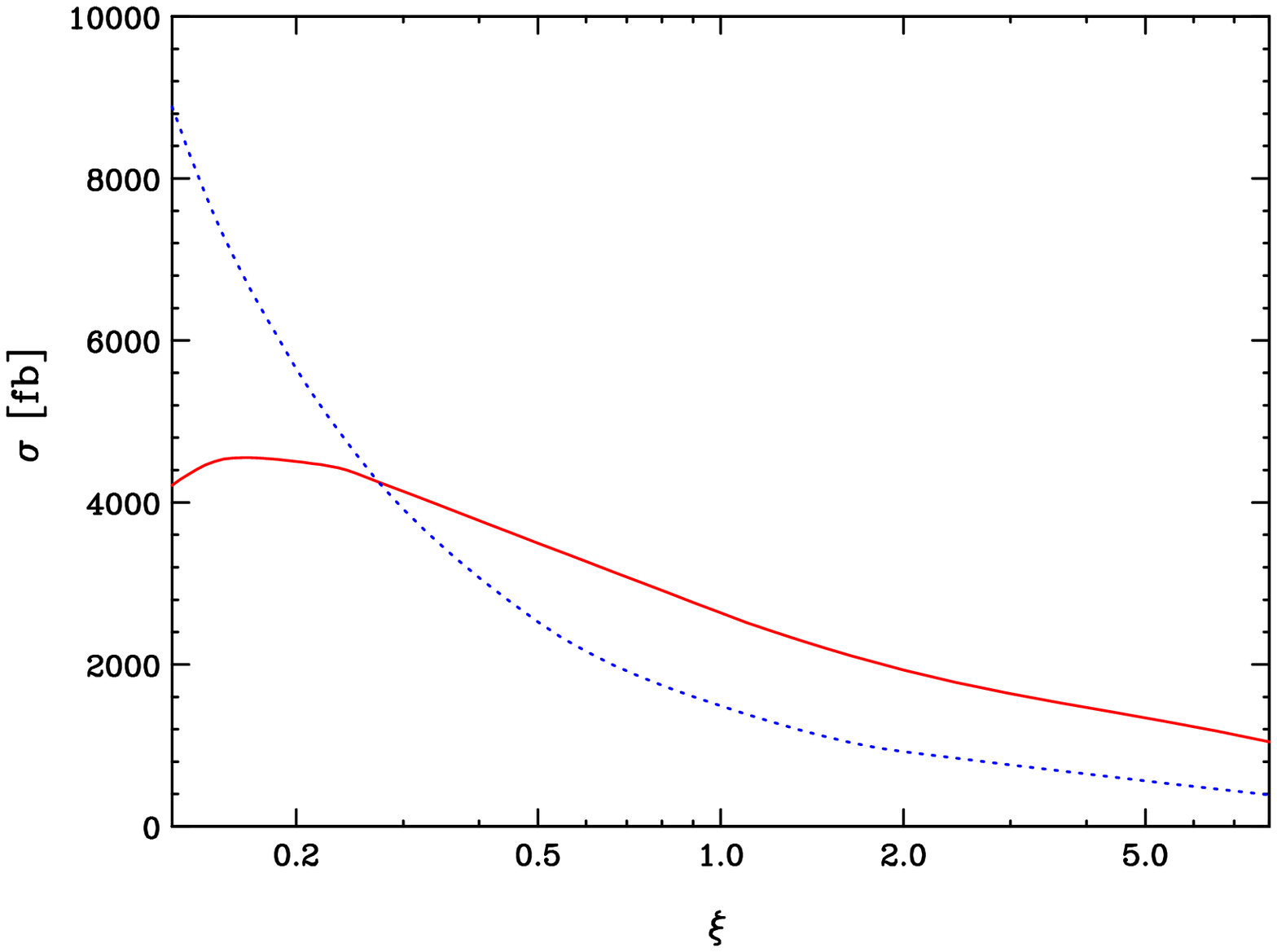}
\includegraphics[width=0.45\textwidth]{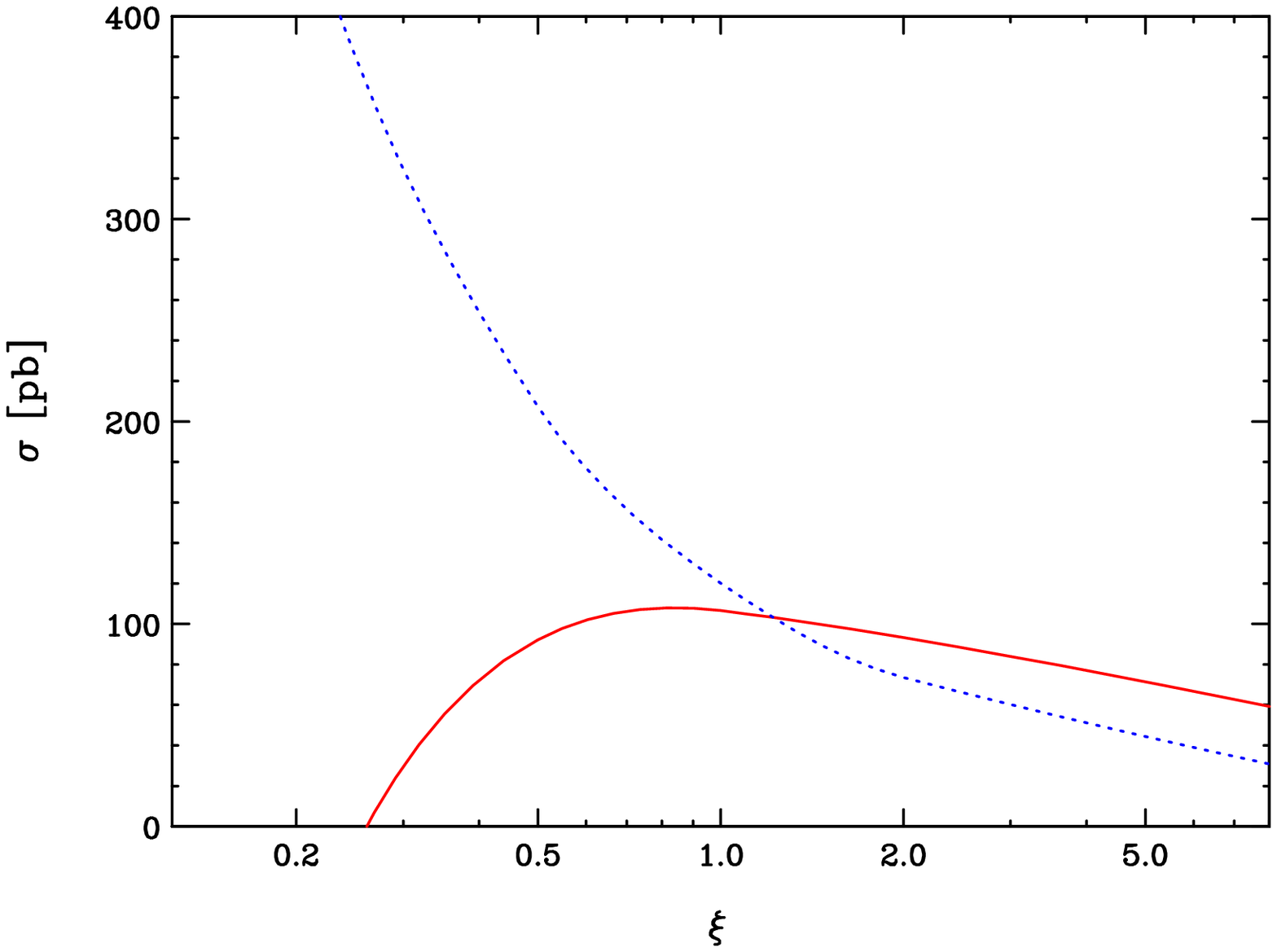} 
\caption{Scale dependence of the total cross section for $pp\rightarrow
  t\bar{t}b\bar{b} + X$ (left panel) and for $pp\rightarrow t\bar{t} jj + X$
  (right panel) at the LHC   with $\mu_R=\mu_F=\xi \cdot \mu_0$ where
  $\mu_0=m_t=172.6$ GeV.   The blue dotted curve corresponds to  the LO
  whereas the red solid to the NLO one.
\label{fig:scales}}
\end{center}
\end{figure}
\begin{figure}
\begin{center}
\includegraphics[width=0.45\textwidth]{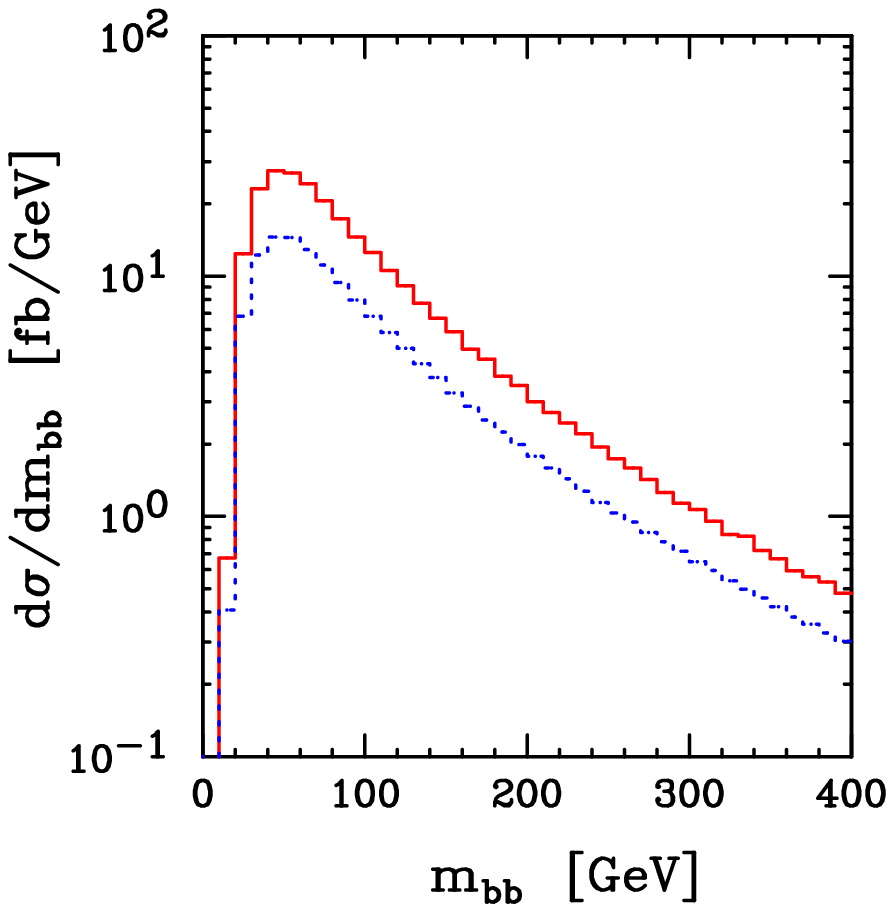}
\includegraphics[width=0.45\textwidth]{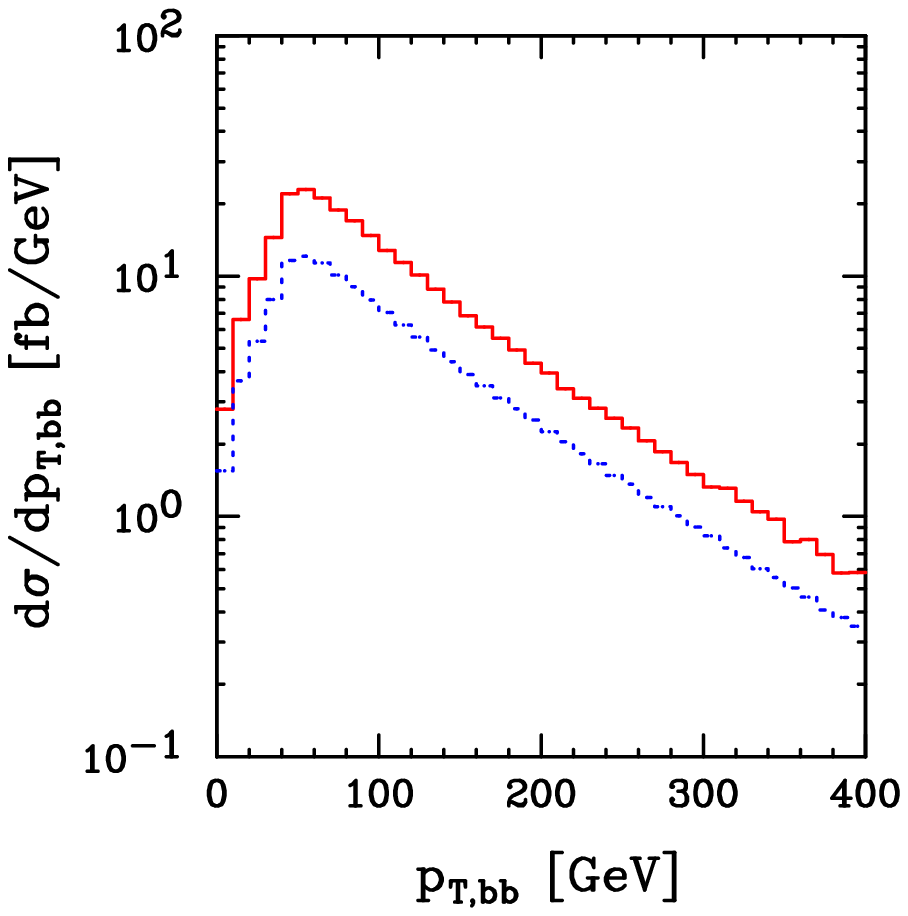} 
\caption{
Distribution of the invariant mass $m_{b\bar{b}}$ (left panel) and
the distribution in the transverse momentum  $p_{T_{b\bar{b}}}$  (right
panel)  of the bottom-anti-bottom  pair for $pp\rightarrow t\bar{t}b\bar{b}
+ X$ at the LHC.  The blue dotted curve corresponds to  the LO  whereas the
red solid to the NLO one.
\label{fig:ttbb}}
\end{center}
\end{figure}
\begin{figure}
\begin{center}
\includegraphics[width=0.45\textwidth]{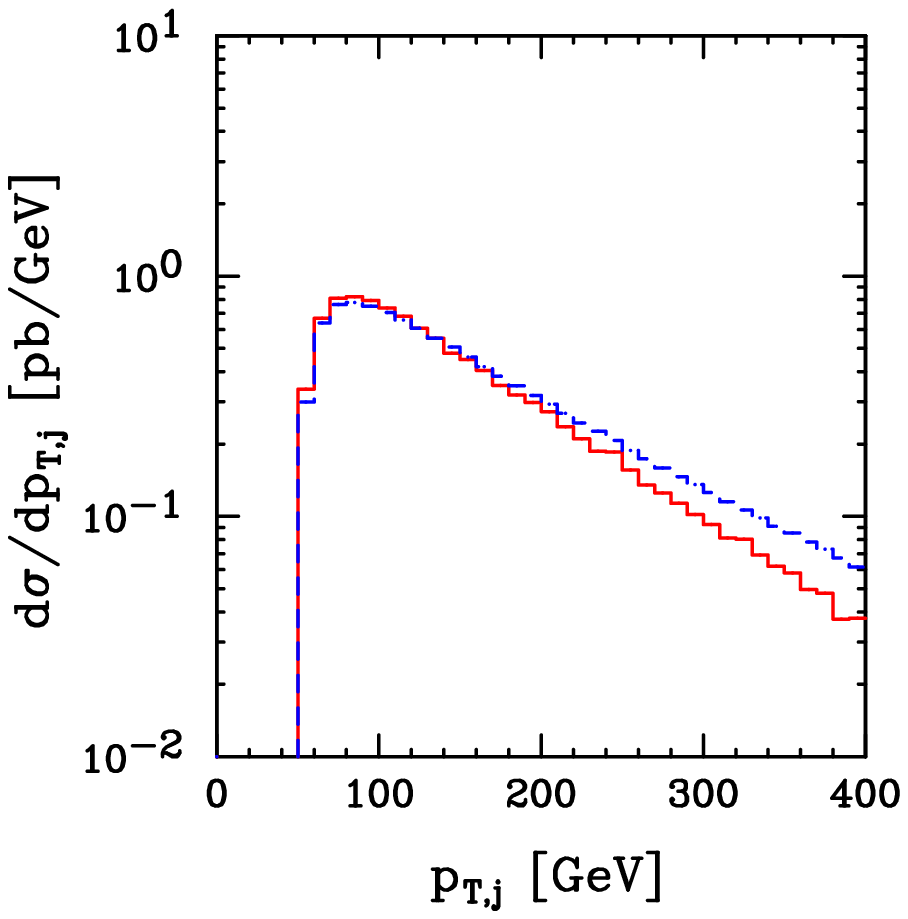}
\includegraphics[width=0.45\textwidth]{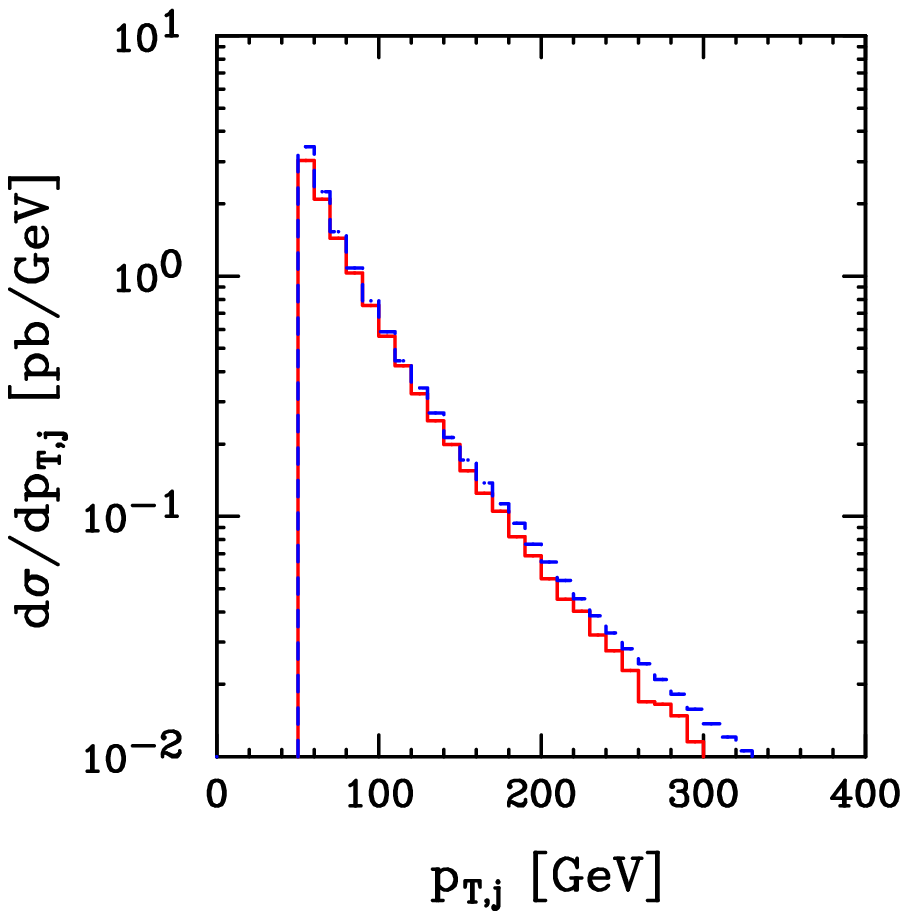} 
\caption{
Distribution in the transverse momentum $p_{T_{j}}$  of the   1st
hardest jet (left panel) and the 2nd hardest jet (right panel)  for
$pp\rightarrow t\bar{t} jj +X$ at the LHC.  The blue dotted curve
corresponds to  the LO  whereas the red solid to the NLO one.
\label{fig:ttjj}}
\end{center}
\end{figure}

\section{Results}

We consider proton-proton collisions at the LHC with a center of mass energy
of $\sqrt{s}=14$ TeV.  The mass of the top quark is set  
to be $m_t=172.6$ GeV. We leave it on-shell with unrestricted 
kinematics. The jets are defined by at
most  two partons using the $k_T$ algorithm with a separation $\Delta R=0.8$,
where $\Delta R=\sqrt{(y_1-y_2)^2+(\phi_1-\phi_2)^2}$,
$y_i=1/2\ln(E_i-p_{i,z})/(E_i+p_{i,z})$ being the rapidity and $\phi_i$ the
azimuthal angle of parton $i$. Moreover, the recombination is only performed
if both partons satisfy $|y_i|<5$ (approximate detector bounds). We further
assume for $t\bar{t}b\bar{b}$  ($t\bar{t}jj$) processes, that the jets are
separated by  $\Delta R=0.8$ $(1.0)$ and have $|y_{\rm{jet}}| < 2.5$ $(4.5)$.  
Their transverse momentum is required to be larger than $20$ $(50)$ GeV 
respectively.  We consistently use the CTEQ6 set of parton distribution
functions, i.e.  we take CTEQ6L1 PDFs with a 1-loop running $\alpha_s$ in LO
and CTEQ6M PDFs with a 2-loop running $\alpha_s$ at NLO.

We begin our presentation of the final results of our analysis with a
discussion of the total cross section. For the central value of the
scale, $\mu_R=\mu_F=\mu_0=m_t$, results for $t\bar{t}b\bar{b}$ 
  production are summarized in Tab.\ref{tab:ttbb} whereas results for 
 $t\bar{t}jj$ production in Tab.\ref{tab:ttjj}.
From the above result one can obtain $K$ factors
\[
K_{pp\rightarrow t\bar{t}b\bar{b}+X}= 1.77\;,
~~~~~~K_{pp\rightarrow t\bar tjj+X} =  0.89\;.
\]
In case of $pp\rightarrow t\bar{t}b\bar{b}+X$ corrections are large of the
order of  $77\%$.  However, they can be reduced substantially, even down to
$-11\%$,   either by applying   additional cuts or by a better choice of 
factorization and renormalization  scales as already suggested by Bredenstein
et al. \cite{Bredenstein:2010rs}.   In case of $pp\rightarrow t\bar{t}jj+X$ we
have obtained negative  corrections of the order of 11\%. In both cases a
dramatic reduction of the scale uncertainty is observed while going from LO to
NLO.  The residual scale uncertainties of the NLO predictions 
for the irreducible background are at the 33\% level, while for 
the reducible background the error obtained by scale
variation is  of the order of  11\%.
The scale dependence of the corrections for both processes is graphically 
presented in Fig.~\ref{fig:scales}. 

While the size of the corrections to the total cross section is
certainly interesting, it is crucial to study the corrections to the
distributions.
In Fig.~\ref{fig:ttbb} the differential distributions for two
observables, namely the invariant mass and transverse momentum
of the two-$b$-jet system are depicted for the $pp\rightarrow t\bar{t}b\bar{b}
  + X$ process. Clearly, the distributions show
the same large corrections, which turn out to be relatively constant
contrary to the quark induced case \cite{Bredenstein:2008zb}.
In Fig.~\ref{fig:ttjj} the transverse momentum distributions of the hardest 
and  second hardest jet are shown for the $pp\rightarrow t\bar{t}jj+ X$ 
process. Distributions demonstrate tiny corrections
up to at least 200 GeV, which means that the size of the corrections
to the cross section is transmitted to the distributions.
On the other hand, strongly altered shapes are  visible
at high $p_T$ especially in case of the first hardest jet. Let us underline
here, that corrections to the high $p_T$ region can only be correctly 
described by higher order calculations and are not altered by soft-collinear 
emissions simulated by parton showers.

\section{Summary}

A brief summary of the calculations of NLO QCD corrections to the 
background processes $pp\rightarrow t\bar{t}b\bar{b} + X$ and
$pp\rightarrow t\bar{t}jj + X$ at the LHC has been presented. They have been 
calculated with the help of the \textsc{Helac-Nlo} system.

The QCD corrections to the integrated
cross  section for the irreducible background   are found to be very large,
changing the LO results by about 77\%.  The distributions show the same large
corrections which are relatively constant.
The residual scale uncertainties of the NLO predictions are at
the 33\% level.
On the other hand, the corrections to the reducible background with respect to
LO are negative and small, reaching 11\%. The error obtained by scale
variation is  of the same order.  The size of the corrections to the cross
section is transmitted to the  distributions at least for the low $p_T$ region.
However, the shapes change appreciably  at high $p_T$.

\acknowledgments

I would like to thank the organizers for  the kind
invitation and very pleasant atmosphere during the conference.  

The work
presented here was  funded by the Initiative and Networking Fund of the
Helmholtz Association,  contract HA-101 (``Physics at the Terascale'') and by
the RTN European Programme MRTN-CT-2006-035505  \textsc{Heptools} - Tools and
Precision Calculations for  Physics Discoveries at Colliders.

\end{document}